\documentclass[allclo]{FBSart}
\usepackage{amsfonts}
\usepackage{amssymb}
\newif\ifpdf
        \ifx\pdfoutput\undefined
        \pdffalse 
        \else
        \pdfoutput=1 
        \pdftrue
        \fi
\ifpdf
        \usepackage[pdftex]{graphicx}
        \DeclareGraphicsExtensions{.pdf, .jpg}
\else
        \usepackage[dvips]{graphicx}
                \DeclareGraphicsExtensions{.eps, .jpg}
\fi

\def\be{\begin{equation}}
\def\ee{\end{equation}}
\def\bea{\begin{eqnarray}}
\def\eea{\end{eqnarray}}
\def\eq#1{(\ref{#1})}

\def\fig#1{Fig.~\ref{#1}}

\title{Level Density of the H\'enon-Heiles System 
       Above the Critical Barrier Energy}
\author{M. Brack\instnr{1,}\thanks{\textit{E-mail address:} 
 matthias.brack@physik.uni-regensburg.de}, J. Kaidel\instnr{1}, 
          P. Winkler\instnr{2}, S. N. Fedotkin\instnr{1,3}}
\instlist{Institut f\"ur Theoretische Physik, Universit\"at Regensburg,
          D-93040 Regensburg, Germany
          \and Department of Physics, University of Nevada
          Reno, NV 89557, USA
          \and Institute for Nuclear Research, 252028 Prospekt Nauki 47,
          Kiev-28, Ukraine}
\runningauthor{M. Brack, S. N. Fedotkin, J. Kaidel and P. Winkler}
\runningtitle{Level Density of the H\'enon-Heiles System 
              Above the Critical Barrier Energy}
\begin{document}

\maketitle
\begin{abstract}
We discuss the coarse-grained level density of the H\'enon-Heiles
system above the barrier energy, where the system is nearly
chaotic. We use periodic orbit theory to approximate its oscillating 
part semiclassically via Gutzwiller's semiclassical trace formula 
(extended by uniform approximations for the contributions of 
bifurcating orbits). Including only a few stable and unstable orbits, 
we reproduce the quantum-mechanical density of states very accurately.
We also present a perturbative calculation of the stabilities of two
infinite series of orbits (R$_n$ and L$_m$), emanating from the shortest 
librating straight-line orbit (A) in a bifurcation cascade just below 
the barrier, which at the barrier have two common asymptotic Lyapunov 
exponents $\chi_{\rm R}$ and $\chi_{\rm L}$.
\end{abstract}

The two-dimensional H\'enon-Heiles (HH) Hamiltonian
\begin{equation}
H_{HH} = T + V_{HH}(x,y) = \frac12\,(p_x^2+p_y^2)+\frac12\,(x^2+y^2)
                           +\alpha\,(x^2 y-y^3\!/3) 
\label{hhh}
\end{equation}
was introduced \cite{hh64} to describe the mean gravitational field 
of a stellar galaxy. It describes an open system in which a particle 
can escape over one of three barriers with critical energy $E_{bar}=
1/6\alpha^2$ and
has meanwhile become a textbook example \cite{ford,gubu,bbook} of a 
system with mixed dynamics reaching from integrable motion (for $E\to 
0$) to nearly fully chaotic motion (for $E\gtrsim E_{bar}$). Scaling 
coordinates and momenta with $\alpha$ causes the classical 
dynamics to depend only on the scaled energy $e=E/E_{bar}=6\alpha^2 E$; 
the barrier energy then lies at $e=1$. 

The Hamiltonian \eq{hhh} has also been used \cite{sadov} to describe the 
nonlinear normal modes of triatomic molecules, such as H$^+_3$, whose 
equilibrium configuration has $D_3$ symmetry. Although this model may no
longer be quantitative for large energies, it can qualitatively 
describe the dissociation of the molecule for $e>1$. 

In this paper we discuss the coarse-grained level density of the HH 
Hamiltonian \eq{hhh} above the barriers, calculated both 
quantum-mechanically and semiclassically using periodic orbit theory. 
Since the potential $V_{HH}$ in \eq{hhh} goes asymptotically to $-\infty$ 
like $-r^3$ ($r^2=x^2+y^2$) in some regions of space, the quantum spectrum 
of \eq{hhh} is strictly speaking continuous. However, for sufficiently 
small $\alpha$ there are quasi-bound states for $E<E_{bar}$ whose widths 
are exponentially small except very near $E_{bar}$. For semiclassical
calculations of the HH level density for $e<1$, we refer to earlier
papers \cite{hhuni,jkmb}.
In \cite{jkpwmb} we have calculated the complex resonance energies 
$E_m-i\Gamma_m$ by the standard method of complex rotation, diagonalizing 
\eq{hhh} in a finite harmonic-oscillator basis. The level density is,
after subtracting the non-resonant part of the continuum, given by
\be
\Delta g(E) = -\frac{1}{\pi}\,\Im \sum_m\frac{1}{E-E_m+i\,\Gamma_{\!m}/2}\,.
\label{dos}
\ee 
We define the {\it coarse-grained} level density by a Gaussian convolution 
of \eq{dos} over an energy range $\gamma$
\bea
\Delta g_{\gamma}(E) 
      = \frac{1}{\gamma\sqrt{\pi}}\int_{-\infty}^\infty
            \Delta g(E')\, e^{-(E-E')^2/\gamma^2}\, {\rm d} E', 
\label{ggam}
\eea
which can be done analytically \cite{jkpwmb}. Its oscillating part, which 
describes the {\it gross-shell structure} in the quantum-mechanical
level density, is then given by
\bea
\delta g_{qm}(E) = \Delta g_{\gamma}(E) - {\widetilde{\Delta g}}(E), 
\label{dgqm}
\eea
where ${\widetilde{\Delta g}}(E)$ is the smooth part of \eq{dos} which
we have extracted by a complex version \cite{jkpwmb} of the numerical
Strutinsky averaging procedure \cite{strut}.

Semiclassically, the quantity $\delta g(E)$ can be approximated by
Gutzwiller's trace formula \cite{gutz}, which for a system with two
degrees of freedom reads
\bea
\delta g_{scl}\left(E\right) = 
\frac{1}{\pi \hbar} \sum_{po} \frac{T_{po}(E)}
{r_{po}\sqrt{\left|{\rm Tr\,M}_{po}(E)-2\right|}}\,
e^{-[\gamma T_{po}(E)/2\hbar]^2} 
\cos\!\left[\frac{S_{po}(E)}{\hbar}-\frac{\pi}{2}\,\sigma_{po}\right]\!\!.\!
\label{dgtr}
\eea
The sum goes over all isolated periodic orbits labeled '$po$', and the
other quantities in \eq{dgtr} are the periods $T_{po}$ and actions 
$S_{po}$, the Maslov indices $\sigma_{po}$ and the repetition numbers 
$r_{po}$ of the periodic orbits. M$_{po}(E)$ is the stability matrix
obtained by linearization of the equations of motion along each periodic 
orbit. The Gaussian factor in \eq{dgtr} is the result of a convolution
analogous to \eq{ggam}; it suppresses the orbits with long periods and
hence yields the gross-shell structure in terms of the shortest periodic 
orbits, hereby eliminating the convergence problem characteristic of 
non-integrable systems \cite{gubu}. This use of the trace formula to
describe gross-shell quantum effects semiclassically has found many 
applications in different fields of physics (including interacting fermion 
systems in the mean-field approximation; see \cite{bbook} for examples).

\begin{figure}\begin{center}
\includegraphics[width=\textwidth]{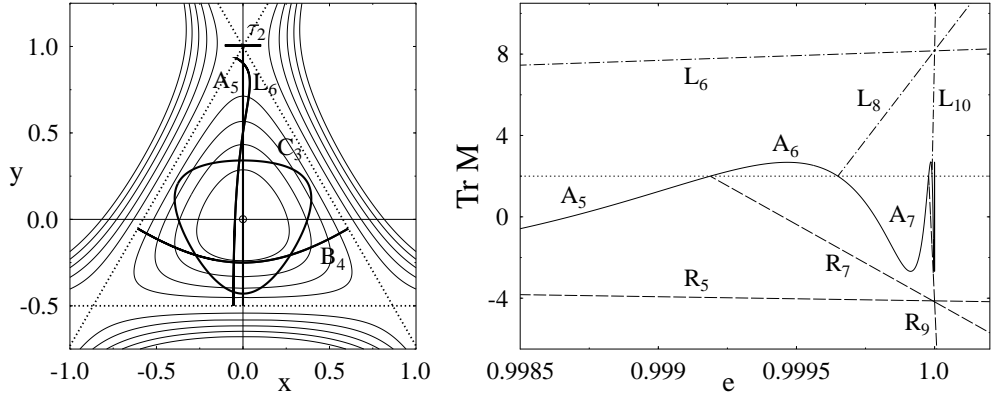}
\caption{{\it Left:} Contours of the HH potential and some of its
  shortest periodic orbits in the $(x,y)$ plane (see text). {\it Right:} Trace
  of the stability matrix of the A orbit and the three pairs of orbits 
  (R$_5$,L$_6$), (R$_7$,L$_8$), (R$_9$,L$_{10}$) bifurcated from it,
  forming the beginnings of the 'HH fans'.\label{combfig}}
\end{center}
\end{figure}

The shortest periodic orbits of the classical HH system \eq{hhh}
have already been extensively studied in earlier papers 
\cite{hhorb,mbgu,lame}. In \cite{jkpwmb} we have calculated all
relevant orbits and their properties from the classical equations of 
motion and computed the quantity $\delta g_{scl}(E)$ in \eq{dgtr}. Some 
of the shortest orbits are shown in the left part of \fig{combfig},
all evaluated at $e=1$ (except for $\tau_2$ which is evaluated at $e=1.1$). 
Note that due to the $D_3$ symmetry of the HH potential,
the orbits A$_5$, B$_4$, $\tau_2$ and L$_6$ (as well as all orbits
R$_n$ and L$_m$ bifurcating from A, see below) have two symmetry partners
obtained by rotations about $\pm 2\pi/3$. The orbit C$_3$ and all 
triplets of R$_m$ orbits have a time reversed partner 
each.\footnote{We use here the nomenclature introduced in \cite{mbgu,lame}, 
where the Maslov indices $\sigma_{po}$ appear as subscripts of the symbols 
(B$_4$, R$_5$, L$_6$, etc.) of the orbits.}

\begin{figure}\begin{center}
\includegraphics[width=\textwidth]{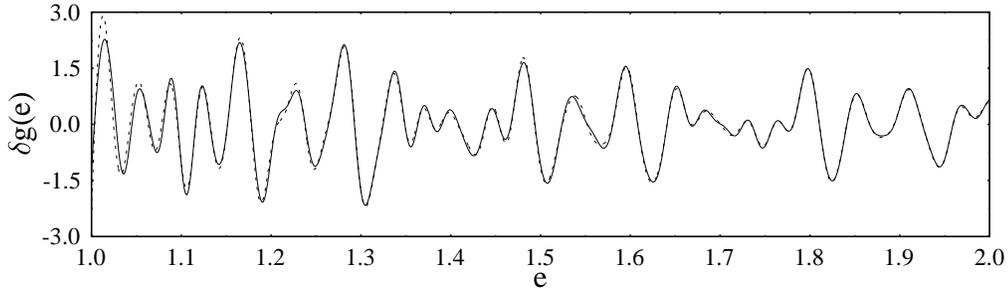}
\caption{Comparison of quantum-mechanical (solid line) and
  semiclassical (dashed line) level density $\delta g(E)$ of the HH potential
  versus scaled energy $e$, coarse grained with Gaussian smoothing
  range $\gamma=0.25$. Only 18 periodic orbits contribute to the
  semiclassical result \cite{jkpwmb}.\label{cont3}}
\end{center}
\end{figure}

In \fig{cont3} we show a comparison of semiclassical \eq{dgtr} with 
quantum-mechanical \eq{dgqm} results, both coarse-grained 
with $\gamma=0.25$ (units such that $\hbar=1$). At this resolution of the 
gross-shell structure, only 18 periodic orbits contribute to the semiclassical
result; for the period-two orbit D$_7$ which is stable up to $e\simeq 1.29$ 
and involves to further orbits (E$_8$, G$_7$) in a codimension-two
bifurcation scenario, we have used the appropriate uniform approximation 
\cite{schom} to avoid the divergence of the trace formula \eq{dgtr} (see 
\cite{jkpwmb} for details). We note that the agreement of semiclassics with 
quantum mechanics is excellent. Only near $e\sim 1$ there is a slight 
discrepancy which is mainly due to some uncertainties in the numerical 
extraction of ${\widetilde{\Delta g}}(E)$. We can conclude that also in the 
continuum region above a threshold, the semiclassical description of quantum 
shell effects in the level density of a classically chaotic system works 
quantitatively.

In view of the importance of the level density close to the critical 
barrier energy $e=1$ for the threshold behaviour of a reaction 
described by the HH model potential, we focus now on a particular 
set of periodic orbits existing at $e=1$. The straight-line librating
orbit A reaches this 
energy with an infinite period after undergoing an infinite cascade 
of bifurcations for $e\to 1$. At these bifurcations, two alternating 
infinite sequences of rotational orbits R$_n$ ($n=5,7,9,\dots$) and 
librating orbits L$_m$ 
($m=6,8,10,\dots$) are born; their bifurcation energies $e_n$ and 
$e_m$ form two geometric progressions converging to $e=1$ with a 
'Feigenbaum constant' $\delta=\exp(2\pi\!/\!\sqrt{3})=37.622367\dots$;
the shapes of these new orbits are self-similar when scaled with 
$\sqrt{\delta}$ in both $x$ and $y$ direction (see \cite{mbgu} 
for details). The stability traces of the first three pairs (R$_n$, 
L$_m$) are shown in the right part of \fig{combfig}. As seen there, 
the curves TrM$(e)$ of these orbits are nearly linear (at least up 
to $e\sim 1.02$) and intersect at $e=1$ approximately at the same 
values for each type (R or L). For large $n$ and $m$, these values 
were found numerically \cite{mbgu} to be TrM$_{{\rm L}_m}(e=1)\sim 
8.183$ and TrM$_{{\rm R}_n}(e=1)\sim -4.183$.  This means that at 
$e=1$, all L orbits have asymptotically the same Lyapunov exponent 
$\chi_{\rm L} \simeq 2.087$, and all R orbits have the same Lyapunov 
exponent $\chi_{\rm R} \simeq 1.368$. Based upon these numerical 
findings, we postulate the following asymptotic behaviour:
\be
{\rm TrM}_{{\rm R}_n,{\rm L}_m}(e) \; \sim \;
           2\mp 6.183 \left(\frac{e-e^*}{1-e^*}\right)
           \quad \hbox{ for } \quad e\to 1\,,\quad n,m\to\infty\,.
\label{hhfan}
\ee
Here $e^*$ are the respective bifurcation energies of the orbits
($e_n$ or $e_m$), and the minus or plus sign is to be associated with 
the R or L orbits, respectively. The curves TrM$_{{\rm R}_n,{\rm L}_m}(e)$ 
thus form two 'fans' spreading out from the values 8.183 and $-4.183$ at 
$e=1$, the first three members of each being shown in the right part of 
\fig{combfig}. In the following we will sketch briefly how the qualitative 
features in \eq{hhfan} of 
these 'HH fans' can be obtained analytically from semiclassical 
perturbation theory. Details will be given in a forthcoming publication 
\cite{fedot}.

The idea is to start from the following 'separable HH' (SHH) Hamiltonian
\be
H_0 = H_{SHH}
    = \frac12\,(p_x^2+p_y^2)+\frac12\,(x^2+y^2)-\frac{\alpha}{3}\,y^3
\label{shh}
\ee
and to include the term $\alpha\,x^2y$ in first-order perturbation
theory. Formally, we multiply it by a small positive number
$\epsilon$ and write $H_{HH} = H_0 + \epsilon H_1$ with 
$H_1=\alpha\,x^2y=u^2v/\alpha^2$,
where $u=\alpha x$ and $v=\alpha y$ are the scaled coordinates.
The Hamiltonian \eq{shh} is integrable; an analytical trace formula
for it has been given in \cite{jkmb}. There is only one saddle at 
$(x,y)=(0,1)$ with energy $e=1$ and one librating A orbit along the 
$y$ axis which undergoes an infinite cascade of bifurcations for 
$e\to 1$. From it, an infinite sequence of rational tori T$_{lk}$ 
bifurcates, where $l$ is their repetition number and $k=2,3,\dots$ 
counts the bifurcations (and the tori). The $v$ motion of the 
primitive A orbit ($l=1$), having $u_A(t)=0$, is given by
\begin{equation}
v_A(t)=v_1+(v_2-v_1) \, {\rm sn}^2(at,q)\,,\quad 
     a=\sqrt{\frac{v_3-v_1}{6}}\,, \quad
     q = \sqrt{\frac{v_2-v_1}{v_3-v_1}}
\label{vA}
\end{equation}
in terms of the Jacobi elliptic function \cite{abro} ${\rm sn}(z,q)$ with 
modulus $q$. In \eq{vA}, $v_1\leq v_2\leq v_3$ are the turning points of 
the motion along the $v$ axis, defined by $V_{HH}(u$=$0,v_i)=e$ ($i=1,2,3$). 
The tori bifurcating at the energies $e_{lk}$ have the same $v$ 
motion as the A orbit: $v_T(t)=v_A(t)$. Their $u$ motion is given by
\bea
u_T(t)=\sqrt{(e-e_{lk})/3}\,\sin(t+\phi)\,, \qquad e\geq e_{lk}\,,
       \qquad \phi \in [0,2\pi)\,.
\label{uT}
\eea
The angle $\phi$ describes the members of the degenerate families of tori. 

According to semiclassical perturbation theory \cite{crpert}, the
actions $S_{lk}$ of the tori are changed in first order of $\epsilon$ by 
\be
\delta_1 S_{lk}(\phi) = - \epsilon \oint_{lk} H_1(u_T(t),v_T(t))\,{\rm d} t
                      = - \frac{\epsilon}{\alpha^2}\int_0^{T_{lk}^{(0)}} 
                          \! u_T^2(t)\,v_T(t)\,{\rm d} t\,,
\label{delS}
\ee
where $T_{lk}^{(0)}=2\pi k$ are the periods of the unperturbed tori 
\cite{jkmb}. The integral in \eq{delS} takes the form 
$\delta_1 S_{lk}(\phi)=A_{lk}+B_{lk}\cos(2\phi)$. In the asymptotic 
limit $e\to 1$, where $q\to 1$ and $T_A \sim {\rm ln}[432/(1-e)]$, 
the coefficients $A_{lk}$ and $B_{lk}$ can be given analytically 
\cite{fedot}. Integrating the phase shift caused by $\delta_1 S_{lk}
(\phi)$ over the angle $\phi$ (i.e., over the torus T$_{lk}$) yields 
a modulation factor \cite{crpert} ${\cal M}_{lk}$
\be
{\cal M}_{lk} = \frac{1}{2\pi}\int_0^{2\pi} e^{\frac{i}{\hbar}
                \delta_1 S_{lk}(\phi)}\,{\rm d} \phi
              = e^{\frac{i}{\hbar}A_{lk}}\, J_0(|B_{lk}|/\hbar)\,,
\label{modfac}
\ee
to be inserted under the sum of tori in the trace formula for the 
unperturbed SHH system given in \cite{jkmb}. Replacing the Bessel 
function in \eq{modfac} by its asymptotic form $J_0(x) \sim \sqrt{2/\pi x} 
\cos(x - \pi/4)$ yields two terms for each torus T$_{lk}$, corresponding
to the two isolated orbits R and L into which it is broken up by the 
perturbation. Reading off their overall amplitudes ${\cal A}_{\rm R,L}$
in the perturbed trace formula and identifying them with their expression 
for isolated orbits given in \eq{dgtr}, i.e.\ equating
\be
{\cal A}_{\rm R,L} = \frac{1}{\pi\hbar}\,
                 \frac{T_{\rm R,L}}{l\sqrt{|{\rm TrM}_{\rm R,L}-2|}}
\ee
uaing the unperturbed periods $T_{lk}^{(0)}$ for $T_{\rm R,L}$,
we can determine the perturbative expression for the stability traces.
For the first repetitions ($l=1$) they become
\be
{\rm TrM}_{{\rm R}_n,{\rm L}_m}(e) \; \sim \;
           2\mp 5.069 \left(\frac{e-e_{1k}}{1-e_{1k}}\right)
           \quad \hbox{ for } \quad e\to 1\,,
\label{fanpert}
\ee
and thus have exactly the same functional form as in \eq{hhfan}.
Here $e_{1k}$ are the bifurcation energies of the primitive
A orbit; $k=2,3,\dots$ labels the pairs of R$_m$ and L$_n$ 
orbits with $m=2k+1$ and $n=2k+2$, and the signs are to be chosen 
as in \eq{hhfan}. In \eq{fanpert} we have put $\epsilon=1$ which is 
justified since even for this value the perturbations $\delta_1 
S_{lk}$ near $e\sim 1$ are sufficiently small.

Although the perturbative result \eq{fanpert} contains a too small
value of the constant 5.069 (instead of 6.183) by 18\%, it explains 
qualitatively correctly the numerical features of the 'HH fans' in 
\eq{hhfan}, in particular the linear intersection of the curves 
TrM$_{\rm R,L}(e)$ at $e=1$ at two values lying symmetrically to 
$\mathrm{TrM}=+2$.

\begin{acknowledge}
We thank A. G. Magner and L. Wiesenfeld for stimulating discussions.
J.K. and S.N.F. acknowledge financial support from the Deutsche 
Forschungsgemeinschaft (DFG) through the graduate college 638 
``Nonlinearity and Nonequilibrium in Condensed Matter''. 
\end{acknowledge}


\begin{thebibliography}{99}

\bibitem{hh64}  M. H\'enon and C. Heiles, Astr.\ J. {\bf 69} (1964) 73.

\bibitem{ford}  G. H. Walker and J. Ford, Phys.\ Rev.\ {\bf 188} (1969) 416.

\bibitem{gubu}  M. C. Gutzwiller: {\it Chaos in Classical and
                Quantum Mechanics} (Springer, New York, 1990).

\bibitem{bbook} M. Brack and R. K. Bhaduri, {\it Semiclassical
                Physics} (2nd edition, Westview Press, Boulder, 2003).

\bibitem{sadov} N. Fulton, J. Tennyson, D. A. Sadovski\'i and
                B. I. Zhilinski\'i, J. Chem.\ Phys.\ {\bf 99} (1993) 906.

\bibitem{hhuni} M. Brack, P. Meier and K. Tanaka, J. Phys.\ A {\bf 32}
                (1999) 331;\\ 
                M. Brack, S. C. Creagh, J. Law, Phys.\ Rev.\ A {\bf 57}
                (1998) 788;\\
                B. Lauritzen and N. D. Whelan, Ann.\ Phys.\ (N. Y.) {\bf 244}
                (1995) 112;\\M. Brack, R. K. Bhaduri, J. Law, Ch. Maier, 
                M. V. N. Murthy, Chaos {\bf 5} (1995) 317; {\it ibid.} 
                (Erratum) {\bf 5} (1995) 707.

\bibitem{jkmb}  J. Kaidel and M. Brack, Phys.\ Rev.\ E {\bf 70} (2004)
                016206, {\it ibid}.\ {\bf 72} (2005) 049903(E).

\bibitem{jkpwmb}J. Kaidel, P. Winkler and M. Brack, Phys.\ Rev.\ E 
                {\bf 70} (2004) 066208.

\bibitem{strut} V. M. Strutinsky, Nucl.\ Phys.\ A {\bf 122} (1968) 1;\\
                M. Brack and H.-C. Pauli, Nucl.\ Phys.\ A {\bf 20} (1973) 401.
                 
\bibitem{gutz}  M. C. Gutzwiller, J. Math.\ Phys.\ {\bf 12} (1971) 343.

\bibitem{hhorb} R. C. Churchill, G. Pecelli and D. L. Rod in:
                {\it Stochastic Behaviour in Classical and Quantum Hamiltonian
                Systems}, eds.\ G. Casati and J. Ford (Springer,
                1979) p.\ 76;\\ K. T. R. Davies, T. E. Huston and M. Baranger, 
                Chaos {\bf 2} (1992) 215;\\ 
                W. M. Vieira and A. M. Ozorio de Almeida, 
                Physica {\bf D 90} (1996) 9.

\bibitem{mbgu}  M. Brack, Foundations of Physics {\bf 31} (2001) 209. 

\bibitem{lame}  M. Brack, M. Mehta and K. Tanaka, 
                J. Phys.\ A {\bf 34} (2001) 8199.  

\bibitem{schom} H. Schomerus, J. Phys.\ A {\bf 31} (1998) 4167.

\bibitem{fedot} S. N. Fedotkin, A. G. Magner and M. Brack, to be published.

\bibitem{abro}  M. Abramowitz and I. A. Stegun: {\it Handbook 
                of Mathematical Functions} 
                (9th printing, Dover, New York, 1970)

\bibitem{crpert}S. C. Creagh, Ann.\ Phys.\ (N. Y.) {\bf 248} (1996) 60.

\end{thebibliography}
\end{document}